# Recurrence is required to capture the representational dynamics of the human visual system

Tim C Kietzmann[1,2,*], Courtney J Spoerer[1], Lynn Sörensen[3],
Radoslaw M Cichy[4], Olaf Hauk[1],
Nikolaus Kriegeskorte[5]

[1]MRC Cognition and Brain Sciences Unit, University of Cambridge
[2]Radboud University, Donders Institute for Brain, Cognition and Behaviour
[3]Psychology Department, University of Amsterdam
[4]Department of Education and Psychology, Freie Universität Berlin
[5]Department of Psychology, Columbia University

* Correspondence to: tim.kietzmann@mrc-cbu.cam.ac.uk
15 Chaucer Road, Cambridge, CB2 7EF, United Kingdom
ORCID identifier: 0000-0001-8076-6062

**Abstract.** The human visual system is an intricate network of brain regions that enables us to recognize the world around us. Despite its abundant lateral and feedback connections, object processing is commonly viewed and studied as a feedforward process. Here, we measure and model the rapid representational dynamics across multiple stages of the human ventral stream using time-resolved brain imaging and deep learning. We observe substantial representational transformations during the first 300 ms of processing within and across ventral-stream regions. Categorical divisions emerge in sequence, cascading forward and in reverse across regions, and Granger causality analysis suggests bidirectional information flow between regions. Finally, recurrent deep neural network models clearly outperform parameter-matched feedforward models in terms of their ability to capture the multi-region cortical dynamics. Targeted virtual cooling experiments on the recurrent deep network models further substantiate the importance of their lateral and top-down connections. These results establish that recurrent models are required to understand information processing in the human ventral stream.

**Significance Statement**

Understanding the computational principles that underlie human vision is a key challenge for neuroscience and could help improve machine vision. Feedforward neural network models process their input through a deep cascade of computations. These models can recognize objects in images and explain aspects of human rapid recognition. However, the human brain contains recurrent connections within and between stages of the cascade, which are missing from the models that dominate both engineering and neuroscience. Here we measure and model the dynamics of human brain activity during visual perception. We compare feedforward and recurrent neural network models and find that only recurrent models can account for the dynamic transformations of representations among multiple regions of visual cortex.

**Introduction**

Vision relies on an intricate network of interconnected cortical regions along the ventral visual pathway (1). Although considerable progress has been made in characterizing the neural selectivity across much of the system, the underlying computations are not well understood. In human neuroscience and corresponding modelling work, insight has often been generated based on time-averaged data and feedforward computational models. However, the primate visual system contains abundant lateral and feedback connections (2). These give rise to recurrent interactions, which are thought to contribute to visual inference (3–13). Understanding the computational mechanisms of human vision therefore requires us to measure and model the rapid representational dynamics across the different regions of the ventral stream. To accomplish this, we here combine magnetoencephalography (MEG), source-based representational dynamics analysis (14, 15), and deep learning. We shed light onto the underlying computations by estimating the emergence of representational distinctions across time and ventral-stream stages and model the data using feedforward and recurrent deep neural network architectures.

**Results**

MEG data were recorded from 15 human participants (306 sensors, two sessions each), while they viewed 92 stimuli from a diverse set of natural object categories (human and non-human faces and bodies, natural and manmade inanimate objects)(16). We focus on three stages of the ventral visual hierarchy, including early- (V1-V3), intermediate- (V4t/LO), and high-level (IT/PHC) visual areas (17). Cortical sources were based on individual-participant reconstructions of the cortical sheet (based on anatomical MRI) and the source signals were computed using minimum-norm estimation (MNE; 18). The resulting data were subjected to representational dynamics analysis (RDA), a time-resolved variant of representational similarity analysis (RSA). For each region and time point, RDA characterizes the representation underlying the stimulus evoked responses by a representational dissimilarity matrix (RDM). This matrix indicates the degree to which different stimuli evoke similar or distinct response patterns in the neural population (Figure 1A). For each region of interest, the temporal sequence of RDMs forms a movie that captures the representational dynamics as a trajectory in a high-dimensional RDM space.

Visual inspection of the RDM movies illustrates the diverse, and highly dynamic nature of the computations along the human ventral stream (Figure 1B, see Supplementary Movie 1 for the whole sequence). RDMs for different regions exhibit distinct representational geometries at identical time points (Figure 1B, columns), reflecting the fact that different ventral stream regions encode visual input based on different features. In addition, representations within each ventral stream region exhibit dynamic changes (Figure 1B, rows), indicating that the intra-area computations, too, undergo substantial transformations as time progresses.

To gain quantitative insight into the neural representations as they vary across time and space, we used linear modelling to break the RDMs down into their component parts. Each RDM was modelled as a nonnegative combination of a set of component RDMs, capturing multiple representational dimensions thought to be prominent in the ventral stream (1). The linear model captures representations that derive from low-level features (Gabor wavelets used in GIST, 19), as well as more abstract distinctions such as animacy (20–22), real-world size (21, 23), and the category of human faces (24). Finally, more fine-grained categorical distinctions were modelled following the categorical structure of the stimulus set (25; see Supplementary Figures 1-3 for further details). The unique contribution of each model component was quantified as the additional variance explained when the component was added to the model explaining the target RDM (26).

This analysis revealed that, as expected, the unique contribution of low-level image features (GIST) emerges early in V1-3 (significant from ≈ 40 ms after stimulus onset, peaking at ≈ 100 ms) and remains substantial and significant throughout the duration of the stimulus (Figure 1C, top left panel). Low-level features were also found to contribute to the early component of the IT/PHC representation, with the onset trailing V1-3 and the peak at a similar latency (≈ 100 ms). However, in contrast to V1-3, the impact of low-level visual features subsequently diminishes in IT/PHC (while remaining significant) as categorical components come to dominate the representational geometry in a staggered sequence. A unique contribution of the face category component emerges next (Figure 1C, bottom left panel) as low level features fade (peaking at ≈ 130 ms in all areas). The rapid onset and strength of the face effect across ROIs is consistent with a special status of faces in the ventral stream (24, 27). Interestingly, the superordinate division of animacy emerges in reverse cascade (Figure 1C, top right panel): It first appears as a prominent peak in IT/PHC (onset: ≈ 140 ms, peak: ≈ 160 ms), vanishes completely (returning to non-significance at ≈ 200 ms), and then appears as a prominent peak in V4t/LO (onset: ≈ 220 ms, peak: ≈ 260 ms), simultaneously resurfacing in IT/PHC, albeit less strongly. Together, these results appear difficult to reconcile with a feedforward-only model. The staggered emergence of representational distinctions (low-level features, faces, animacy) within a given region, the temporary waning of previously prominent divisions (GIST, faces, animacy), and the reverse cascaded emergence of animacy, all occurring while the stimulus is still on (500 ms), suggest highly dynamic recurrent computations.

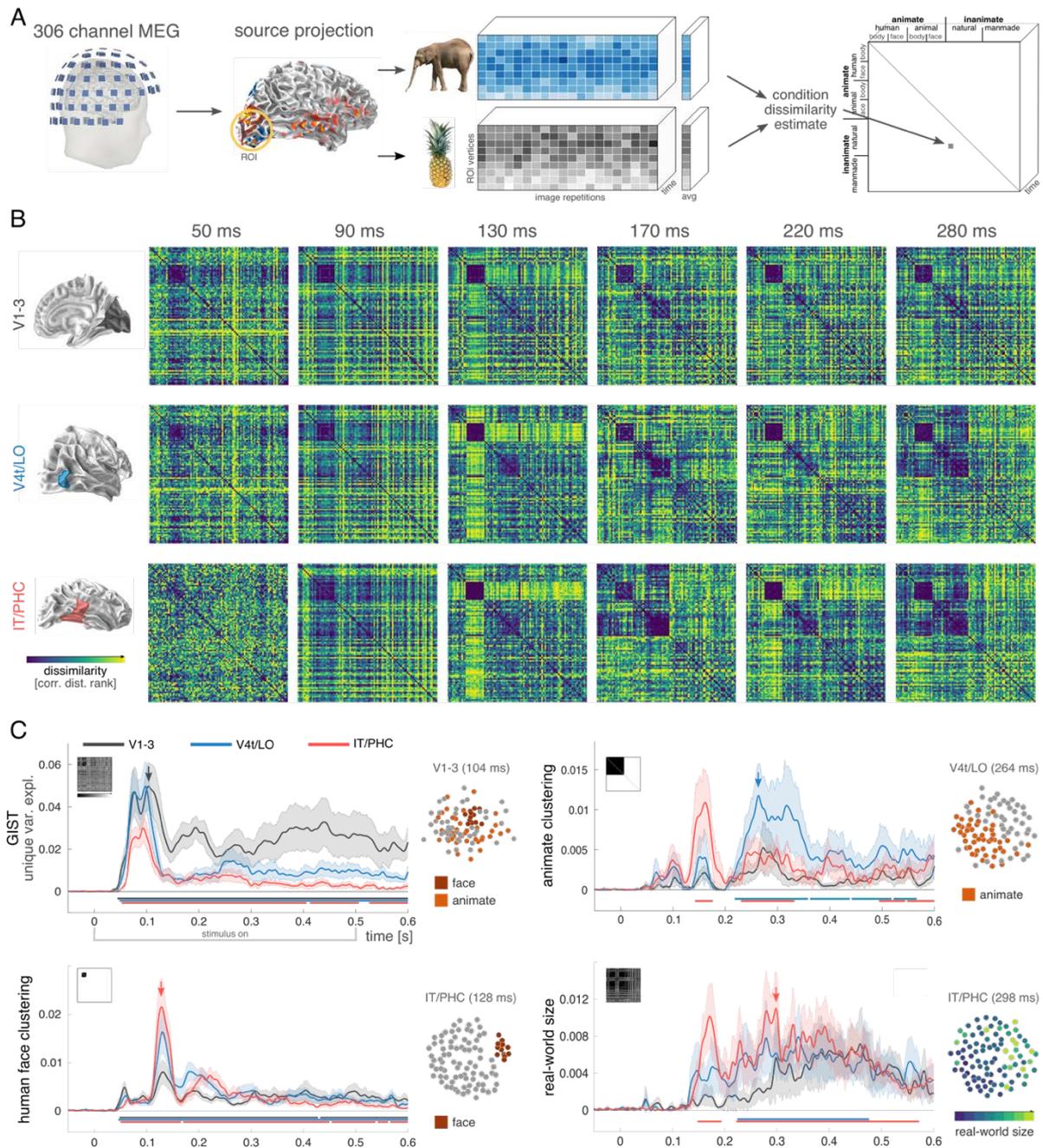

**Figure 1. Representational dynamics analysis (RDA) reveals how feature selectivity emerges over time along distinct ventral stream regions.** (A) RDA pipeline to extract source-space RDM movies. (B) Participant-averaged RDMs (ranked) for regions V1-3, V4t/LO, and IT/PHC at selected time-points. All ROIs exhibit distinctive multi-stage representational trajectories. (C) Linear modelling of the temporal evolution of feature selectivity reveals a staggered emergence of representational distinctions within and across ventral stream ROIs (black curve = V1-3, blue = V4t/LO, red = IT/PHC). Horizontal bars indicate time points with effects significantly exceeding pre-stimulus baseline (nonparametric cluster-correction; cluster inclusion and significance level $p<0.05$). Standard error across participants shown as shaded area. Representational geometries at selected time points and ROIs are visualized in 2D using multidimensional scaling (MDS) to visualize effect magnitude.

As an additional test for recurrent interactions across the ventral stream ROIs, we performed bottom-up and top-down Granger-causality analysis, testing in how far the past of a source ROI can improve predictions of the RDMs observed in a target ROI (Figure 2; see Methods for details). Compatible with a feedforward flow of information, Granger causality was found to be significantly above baseline from V1-3 to V4t/LO and from V4t/LO to IT/PHC, emerging around 70 ms after stimulus onset in each case. In addition, Granger causality was significant in the feedback direction, emerging more gradually with a peak just past 110 ms for V4t/LO to V1-3, and peaks around 140 ms and 260 ms for IT/PHC to V4t/LO. While the current Granger causality model did not include common input to source and target regions, the bidirectional influence observed is difficult to reconcile with confounding input at differential delays from a third lower-level region.

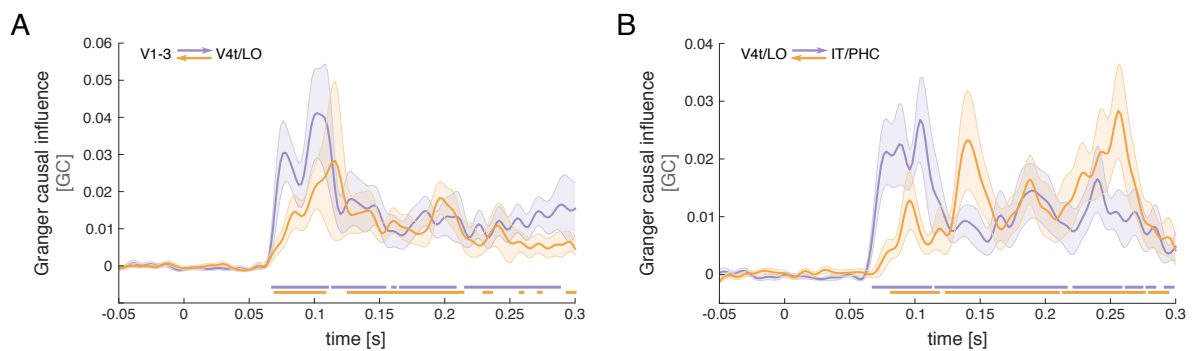

**Figure 2. RSA Granger causality analysis was performed to estimate information flow between ventral stream areas.** (A) Feedforward (purple) and feedback (orange) direction of Granger causal influence between early- and intermediate ROIs, and (B) effects between intermediate- and high-level ROIs. Horizontal bars indicate time points with causal interactions exceeding effects during pre-stimulus baseline (FDR corrected at $q<0.05$). Data are shown baseline corrected.

Our analyses thus far reveal rich representational dynamics within ROIs, as well as bidirectional information flow between ventral stream regions. These suggest a prominent role of recurrence in computations along the ventral visual pathway. We next tested this hypothesis more directly using deep learning (28–31) to obtain image-computable models of brain information processing. We trained different deep neural network (DNN) architectures to mirror the time-varying representations of all ventral-stream areas (Figure 3A). The trained models were then compared in terms of their ability to predict held-out MEG data. This modelling approach offers a direct test for the representational capacity of a given network architecture and thereby helps distinguish between competing hypotheses about the underlying computations. Two classes of convolutional neural network architecture were tested: feedforward and recurrent. Standard feedforward architectures, including commonly used off-the-shelf pre-trained DNNs, do not express any dynamics, as each layer produces a single activation vector which is passed on to the next. To maximize the potential for complex dynamics within the feedforward framework, we therefore allowed units to ramp-up their activity over time. This was achieved via self-connections, whose weights were optimized along with the other parameters to best match the MEG data. Ramping feedforward models can exhibit complex dynamics, capturing for example the way neurons integrate incoming signals and accumulate evidence. While this technically constitutes a recurrent architecture, it does not enable lateral and top-down message passing. Ramping feedforward models include pure feedforward DNNs as a special case and therefore provide a more conservative control in testing the hypothesis of recurrent computation in the ventral stream. The recurrent models included bottom-up, lateral, and top-down connections (BLT; 6), i.e. local recurrence within network layers/regions (L) and bi-directional connections across layers (B and T). The latter enabled us to model feedback between ventral stream ROIs, expanding on previous work investigating the effects of recurrence within a given region while restricting cross-regional information flow to the feedforward direction (32, 33). Importantly, a meaningful comparison between recurrent and feedforward architectures requires the control of as many architectural differences as possible. These include, among others, the number of layers, feature maps, and the total number of network parameters, all of which can affect a network's ability to fit to the data presented. To control for the additional parameters introduced by lateral and top-down connections in the recurrent networks we varied the kernel sizes in two ramping feedforward models ($B_{K11}$ – kernel size 11, and $B_{K9}$ – kernel size 9; for a similar approach see (34, 35)). This allowed us to approximately match the number of parameters across network architectures (see Methods for details), and hence directly test for the effects of added recurrence.

To test the different network architectures for their capacity to mirror the human ventral stream dynamics we introduced a novel deep learning objective that uses the RDM data of the three ventral-stream ROIs as targets for the representations in separate network layers. Using backpropagation to learn the network weights, this objective optimizes each model to best predict the MEG RDM movies (dynamic representational distance learning (dRDL), see 36 and Methods for details). The model time steps were set up to mirror a 10 ms delay from one target ROI to the next (Figure 3A), in line with lower bound estimates for information transfer across ventral-stream regions (37). To avoid overfitting to the 92 experimental stimuli, an independent set of 141,000 novel images originating from the same object categories was used for network training (Supplemental Figure 4). Each trained network was tested on the previously unseen experimental stimuli, and the fit between the network RDM movies and the MEG RDM movies was estimated by cross-validation (see Supplemental Figure 5 and Supplemental Movie 2-6 for a direct comparison of model and ventral-stream RDM movies).

We first compared the trained DNNs to the ventral stream ROI dynamics in terms of the average representational distance across all stimulus pairs as it varies across time (Figure 3B). While ramping feedforward networks exhibit complex representational dynamics, their average representational distances did not closely follow the empirical data, especially in higher-level ventral-stream regions (average-distance trajectory correlations with held-out data: 0.83, 0.59, 0.47 for V1-3, V4t/LO, and IT/PHC, respectively). In contrast, recurrent DNNs almost perfectly matched the average distances of all ventral-stream ROIs (average-distance trajectory correlations: 0.95, 0.93, 0.97 for V1-3, V4t/LO, and IT/PHC, respectively; significantly outperforming ramping feedforward models for all ROIs and cross-validation splits at p<0.0001 using Hittner's *r* to *z* procedure (38, 39)), despite being tested on a new set of stimuli and compared against held-out MEG data. For a more detailed comparison of the patterns of representational distances, we next evaluated how well the model RDM movies matched the ventral stream data frame by frame. For each time point, we computed the correlation between the RDM of the corresponding model layer and the ventral-stream RDM. These correlations were averaged across time to yield a summary statistic (Figure 3C, see Supplemental Figure 6 for the full time-courses). For each ventral-stream area, the recurrent model significantly outperformed the ramping feedforward models (Wilcoxon signed-rank test, p<0.005 in all cases). The recurrent models also outperformed a layer-based readout from commonly used computer vision models Alexnet and VGG16 (Supplemental Figure 7). We also tested the DNNs, trained on the time-varying MEG data, for their ability to predict temporally static functional magnetic resonance imaging (fMRI) data acquired from the same participants and ROIs. Again, recurrent models provided a significantly better prediction to ramping feedforward models (Figure 3D; p<0.001 for V1-3, p<0.05 for V4t-LO, p<0.001 for IT/PHC; see Methods and Supplementary Figure 8 for details). Finally, the recurrent architectures also outperformed the ramping feedforward models in terms of classification performance on the held-out image test set by a large margin (top-1accuracy ~64% for the ramping feedforward models ($B_{K9}$, $B_{K11}$), and 73.9% for the recurrent models). These results add to the growing body of literature suggesting that the performance computer vision applications can be improved by integrating neuroscientific computational principles, such as recurrence (5, 34, 35, 40, 41), and neuroscientific data (42).

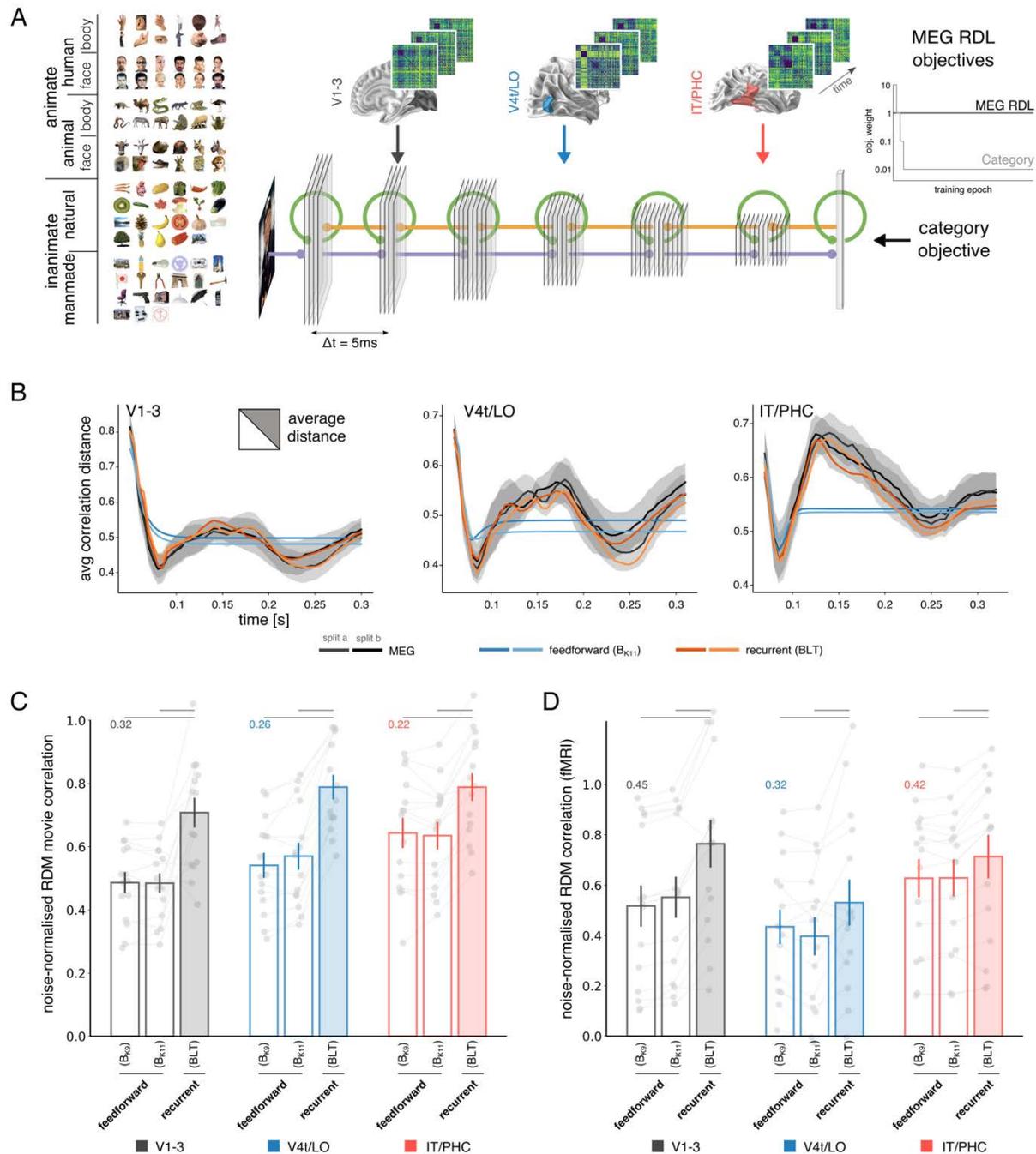

**Figure 3. Deep neural network modelling of ventral stream representational dynamics.** (A) The RDM movies of all three ventral stream regions were used as a time-varying deep learning objectives targeting separate DNN layers, together with a time-decaying category objective at the readout. Each artificial network thereby attempts to simultaneously capture the representational dynamics of all ventral stream areas. (B)

Development of the average pattern distance across time. MEG data are shown together with the results of ramping feedforward, and recurrent DNNs. (C) Average frame-by-frame RDM correlation between model and brain. Correlations estimated on separate data from individual participants, shown as gray dots. Data are normalized by the predictive performance of the MEG RDM movies used for training (normalization factor shown for each region at the level of 1.0). For all ROIs (black = V1-3, blue = V4t/LO, red = IT/PHC), recurrent networks significantly outperform ramping feedforward architectures (significance indicated by gray horizontal lines above). (D) Cross-validated predictive performance of different DNN architectures trained on the MEG data when tested against fMRI RDMs, acquired from the same participants and ROIs. Correlations were noise-normalized using the respective lower bound of the noise ceiling. For all regions, recurrent networks significantly outperform ramping feedforward architectures.

To better understand the connectivity within the recurrent networks, we performed virtual cooling experiments in which we increasingly deactivated specific connection types (lateral and top-down) in distinct network layers. We then tested the resulting DNNs for their ability to (a) perform object classification and (b) model human ventral stream dynamics. For object classification, we observed that lateral and top-down connections in lower layers had a stronger impact on performance, with strong effects resulting from cooling top-down connections into the network layer modelling V1-3 (Figure 4A). For predicting ventral stream dynamics, we again found that both connection types were of importance, although the success of higher-level ventral stream predictions was less reliant on top-down network connections (Figure 4B).

**Conclusions**

Our analyses of the RDM dynamics, Granger causality between regions, and DNN models all consistently show that human ventral-stream dynamics arise from recurrent message passing, which, among other computational functions, may facilitate recognition under challenging conditions (6, 32, 33, 35). The combination of source-based MEG representational dynamics analysis and recurrent deep neural network models opens new horizons for investigation of information processing in the human brain, as well as for novel engineering applications that incorporate neural data into machine learning pipelines.

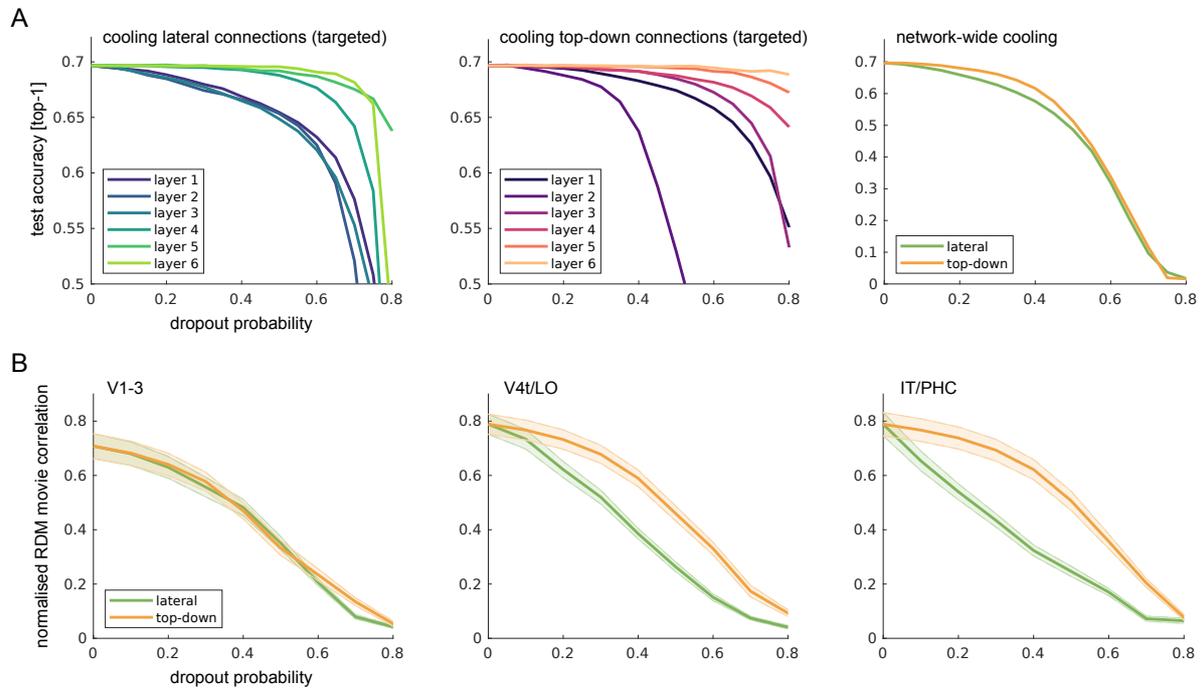

**Figure 4. DNN cooling studies.** (A) Virtual cooling experiments allow for the specific targeting and deactivation of input connections into distinct network layers. The effects of lateral (left) and top-down (middle) computations on object categorization performance vary across network depth, with stronger effects observed while deactivating recurrence in earlier layers. Applied to the whole network (right), the cooling of lateral and top down connections have comparable effects, with perhaps stronger reliance on lateral connectivity. (B) Targeting specific connection types throughout the network reveals the importance of lateral and top-down network connections for modelling human ventral stream dynamics. Later network layers are increasingly robust to the cooling of top-down input.

**Materials and Methods**

## MEG data acquisition, pre-processing and source reconstruction

**Experimental setup**

Data collection procedures and experimental design were described in detail previously (16). MEG data from 16 right-handed participants (10 females, mean age 25.87 years; std = 5.38) were recorded. MEG source reconstruction analyses were performed for a subset of 15 participants for whom additional structural and functional MRI data were acquired. All participants had normal or corrected-to-normal vision and gave written informed consent in each experimental session (two MEG, 1 fMRI for each participant). The study was approved by the Institutional Review Board of the Massachusetts Institute of Technology and conducted according to the Declaration of Helsinki.

During the experiment, participants were shown 92 different objects. This stimulus set was used across multiple studies and laboratories to collect human fMRI (20), and MEG data (16, 43), human perceptual similarity judgments (44), macaque single cell data (27), and was used in previous investigations of deep neural network models (25, 45). The stimulus set therefore allows for comparisons across modalities, species, and recording sites. Furthermore, it includes a large variety of object categories, allowing for a more complete characterization of population responses in the human visual cortex, compared to less diverse sets. It includes depictions of 12 human body parts, 12 human faces, 12 animal bodies, 12 animal heads, 23 natural objects, and 21 artificial/manmade objects.

Each participant completed two experimental MEG sessions. Stimuli were presented on a grey background (2.9 degrees of visual angle, 500 ms stimulus duration), overlaid with a dark grey fixation cross (trial onset asynchrony (TOA) of 1.5 or 2 s). Participants were asked to indicate via button press and eye blink whenever they noticed the appearance of a paper clip. These target trials, occurring randomly every 3-5 trials, were excluded from further analyses. Each session consisted of 10-14 runs, and each stimulus was presented twice in a given run.

**MEG data acquisition and pre-processing**

Data were acquired from 306 MEG channels (102 magnetometers, 204 planar gradiometers) using an Elekta Neuromag TRIUX system (Elekta, Stockholm). The raw data, sampled at 1 kHz, were bandpass filtered between 0.03 and 330Hz, cleaned using spatiotemporal filtering (maxfilter software, Elekta, Stockholm), and subsequently downsampled to 500Hz. Trials were baseline-corrected using a time-window of 100 ms before stimulus onset. For each participant and session, flat sensors and sensors exhibiting excessive noise (defined as baseline variance exceeding a z-threshold of ±3, z-scores computed over the distribution of all sensors of a given type) were removed from further analyses. On average, 2.67 gradiometers (*std* = 1.79) and 0.67 magnetometers (*std* = 1.06) were excluded. Trials with excessive noise were discarded by means of the autoreject toolbox (46). After cleaning, an average of 26.08 (*range* 16 - 35) repetitions per stimulus, participant, and session entered subsequent analyses.

**MEG source reconstruction**

*Source reconstructions*
Source reconstructions were performed using MNE (18), as implemented in the MNE python toolbox (47). Volume conduction estimates were based on participant individual structural T1 scans, using single layer boundary element models (BEMs). BEMs were based on the inner skull boundary (extracted via fieldtrip (48) due to poor reconstruction results from the FreeSurfer (49) watershed algorithm used in MNE python). The source space comprised 10242 source points per hemisphere, positioned along the grey/white boundary, as estimated via FreeSurfer. Source orientations were defined as surface normals with a loose orientation constraint. MEG/MRI alignment was performed based on fiducials and digitizer points along the head surface (iterative closest point procedure after initial alignment based on fiducials). The sensor noise covariance matrix was estimated from the baseline period (-0.1 – 0 s with respect to stimulus onset) and regularized according to the Ledoit-Wolf procedure (50). Source activations were projected onto the surface normal, yielding one activation estimate per point in source space and time.

*Regions of interest*
Three regions of interest (ROIs) were defined along the ventral visual stream, covering early (V1-3), intermediate (V4t/LO1-3) and downstream, high-level visual areas (IT/PHC, consisting of TE1-2p, FFC, VVC, VMV2-3, PHA1-3). ROIs were defined comparably large and spatially distinct to maximize SNR while limiting cross-talk. A potential separation of the early ROI into multiple smaller visual areas is complicated by the small stimulus size (2.9 degree visual angle), preventing a clear attribution of activity near the foveal confluence.

Each ROI was derived from a multi-modal atlas of cerebral cortex, which provides the underlying parcellation (17). The atlas annotation files were converted to fsaverage coordinates (51) and from there mapped to each individual participant via spherical averaging.

**MEG Representational dynamics analysis**

We used a time-resolved extension of representational similarity analysis (RSA) (14) to gain insights into the representational transformations of the visual inputs across time for all three regions of interest. The central element of RSA are representational dissimilarity matrices (RDMs), which characterize how a given ROI distinguishes between experimental conditions. A small distance between a pair of conditions implies a similar neural response, whereas large distances imply that the region treats the two stimulus conditions separately. RDMs thereby equate to representational geometries, which define the spatial relationship of experimental conditions in the underlying activation space. To get a better understanding of the organizational principles underlying a given RDM, computational and categorical models can be used to predict (condition relative) empirical distances. Temporal sequences of RDMs across multiple ROIs can furthermore be used to test for effects of granger causality; i.e. the transfer of representational organizations between ROIs.

**RDM extraction**

To compute temporally resolved RDM movies from MEG source data, we first extracted a single multivariate source time-series for each condition by averaging across repetitions. RDMs were then computed by estimating the pattern distance between all combinations of conditions using correlation distance (1-Pearson correlation). One RDM was computed for each time point, yielding a temporally changing RDM movie (size: n_objects x n_objects x n_timepoints). RDM movies were computed for each participant, ROI, hemisphere, and session separately. We then averaged the RDM movies across hemispheres and sessions, yielding one RDM movie for each ROI and participant. As RDMs are diagonally symmetric, only the upper triangles of the RDM movies were used for subsequent analyses. For visualization purposes, all shown RDMs are rank-transformed. All analyses were performed on the non-transformed data.

**Model fitting and statistics**

To better understand and quantitatively assess representational transformations across time, we modelled the RDM movies of each participant and ROI using a hierarchical GLM. The overall idea of RDM modelling is to define a set of external computational/categorical models, each predicting distinct condition-specific distances, which are then combined to explain the observed empirical distances. These predictors are not necessarily orthogonal, and therefore the actual contribution of each predictor to the overall variance explained can be ambiguous. To solve this issue, we here compute unique variance explained of each model predictor by subtracting the total variance explained of the reduced GLM (excluding the predictor of interest) from the total variance explained by the full GLM. This procedure was followed for each model predictor, participant, ROI, and time-point.

To find the optimal weights for the linear combination of model predictors, we used a non-negative least squares approach. The predictions of four main and 10 additional control predictors were investigated. The resulting fourteen model predictors were standardized before entering the GLM. The main predictors included animate-, and face-clustering, low-level GIST predictions, and representational geometries resulting from organizations based on the real-world size (23). Beyond these four, additional predictors were included, which mirror the categorical structure of the stimulus space: inanimate-, human-, animal-, face- (monkey, inter-species), body (human, monkey), natural- and artificial object clustering. Finally, a constant term was included in the GLM model. Following the GLM modelling approach described above, we obtained unique variance traces across time for each participant, GLM predictor and ROI. Predictor-specific statistical tests were performed across participants for each ROI and time-point.

To establish whether the unique variance explained by a model predictor exceeded the expected increase due to the addition of a free parameter to the GLM, we tested the unique variance observed at each time point against the average increase during the pre-stimulus baseline period. To control for multiple comparisons across time, a non-parametric cluster test was used max-sum test statistic, computed on a paired, one-sided t-statistic (one-sided because effects of interest are strictly larger than the effects observed during baseline); cluster inclusion criterion of $p<0.05$)(52). The statistical baseline period was defined as the 50 ms time-window directly prior to stimulus onset. The first 600 ms of stimulus processing were included in the analyses. Statistical comparisons were performed on the unsmoothed signal. To aid visibility, unique variance curves were low-pass filtered at 80 Hz (Butterworth IIR filter; order 6) prior to plotting.

**RSA Granger analysis**

To investigate the possibility of information transfer between ROIs, we performed a Granger causality analysis on the basis of the RDM movies (53). That is, we asked whether the current RDM of a target ROI could be explained by the past RDMs of a source ROI, beyond the explanation offered by the past of the target ROI itself. As for the model predictions above, this was also implemented by a hierarchical GLM approach (again using non-negative least squares). We first used the past RDMs of the target ROI itself to explain the current RDM, and then tested in how far the addition of the past RDMs from a source RDM would add to the variance explained. Granger causal influence was defined as $GC=\ln(U_{reduced}/U_{full})$ (U = unexplained variance by the reduced and full model, respectively, 54). Again, the inclusion of additional predictors, and therefore free parameters, can by itself lead to an increase in the variance explained. We therefore used the average increase in variance explained during a pre-stimulus tine window (50 ms prior to stimulus onset) as baseline for statistical comparisons. For each pair of adjacent ROIs (V1-3 and V4t/LO1-3, as well as V4t/LO1-3 and IT/PHC), we tested both directions of Granger causality, using the standardized RDMs of each ROI once as source and once as target. To predict the RDM data at time point $t$, we used a 100 ms time-window of $t$-120 ms to $t$-20 ms. To test for effects of Granger causality across time, we performed above analysis separately for each time-point within the first 300 ms post stimulus onset. To correct for multiple comparisons, we performed an FDR correction (p<0.05) for all tested time-points tested for the two source ROIs. Statistical comparisons were performed on the unsmoothed signal. To aid visibility, unique variance curves were low-pass filtered at 80 Hz (Butterworth IIR filter; order 6) prior to plotting.

**Noise ceiling estimates**

We computed the upper and lower bounds of the signal noise ceiling for each region of interest and time point. We computed the lower bound for each participant as the predictive performance of the grand average of all other participants. The upper bound was computed by using the grand average of all participants (55). The latter is overfitted to the respective group of participants, as each individual participant's data is included in the grand average prediction. This renders the upper bound a true ceiling for model predictive performance. As we used nonnegative least squares in the linear modelling analysis, we used the same analysis pipeline to compute the variance explained by the respective grand average data. We report the participant averaged noise ceiling in Supplemental Figure 3.

**FMRI data acquisition and analyses**

FMRI data were collected for 15 participants. Stimuli were presented once per run, participants completed between 10 and 14 runs each. Each run contained additional 30 randomly timed null trials without stimulus presentation. During these trials, participants had the task to report a short (100ms) change in the luminance of the fixation cross via button-press. fMRI experimental trials had a TOA of 3 seconds (6 s in presence of a null trial). For further acquisition details, please see (16). Pre-processing was performed using SPM8 (http://www.fil.ion.ucl.ac.uk/spm/). Functional data were spatially realigned, slice-time corrected and co-registered to the participant-individual T1 structural image. Data were then modelled using a general linear model (GLM), which included movement parameters as nuisance terms. GLM parameter estimates for each condition/stimulus were contrasted against an explicit baseline to yield a t-value for each voxel and condition. The 500 most strongly activated voxels were included in subsequent analyses.

Regions of interest were defined in alignment with the MEG ROIs. The corresponding ROI masks were defined on the individual surface and projected into the functional volume using freesurfer (49). To characterize the representational geometry of a given ROI, the activation patterns (t-values) were extracted for all possible pairs of stimuli, and the pattern distances were computed based on 1-Pearson correlation, in line with the distance measures used in the MEG data.

**RCNN model predictions of fMRI data**

RCNN models, originally fitted to the MEG data, were used to predict the temporally smooth fMRI representational similarities. Since the RCNN models predict temporal sequences of RDMs for each ROI, the time-points of a given layer were linearly combined to obtain a single RDM prediction for the fMRI data. Network layers were chosen for each fMRI ROI to match the corresponding MEG ROI used during training.

The linear weights for the individual time-points were computed using non-negative least-squares, fitting to the average RDM of a given ROI based on the data of N-1 participants. The resulting reconstruction was then used to predict the RDM of the left-out participant. The goodness of fit of this cross-validated prediction was determined by correlating the upper triangles of the two RDMs. Prediction accuracies were statistically compared using random effects test across participants (non-parametric Wilcoxon signed rank test).

# **Neural network models**

We modelled the observed MEG RDM movies with convolutional neural networks implemented using TensorFlow (56). Two specific architectures were tested, feedforward networks, where bottom-up connections dominate (termed 'B' for bottom-up hereafter), and a recurrent network, with bottom-up, lateral and top-down connections (BLT; 6). Feedforward and recurrent models were matched to have approximately the same number of parameters. To enable feedforward networks to exhibit non-trivial dynamics, we allowed the networks to learn to ramp-up the activity of their units over time.

**Training data sets**

Networks were trained using representational distance learning (RDL; 36) to predict the time-varying representational dynamics in the ventral stream up to 300ms after stimulus onset. To train the networks with RDL we collected a data set of 141,000 images – RDL61. This data set consists of 61 categories derived from the 92 images that were used in the human imaging experiments. For each category in the experimental stimulus set, a set of natural images were obtained and subdivided into a training set and a validation set.

**Image pre-processing**

During network training, each image underwent a series of pre-processing steps before being passed to the network. Firstly, a crop was randomly sampled from the image that covered at least a third of the of the image area with an aspect ratio in the range 0.9-1.1 (specified as the ratio width/height). The image was then randomly flipped along the vertical axis, and small random distortions were applied to the brightness, saturation and contrast. Finally, the image was resized to 96×96 pixels.

**Cross-validation**

To avoid overfitting, we cross-validated the networks with respect to both, the input images and the MEG data. Firstly, all of the network responses were analyzed using the same 92 stimuli that were shown to the human participants. These images are both independent and visually dissimilar (showing only a single object on a grey background) from the natural images used to train the networks.

Secondly, networks were evaluated against MEG data that was held out from the model fitting procedure. This was accomplished by assigning single-session data for each subject to one of two splits. Networks were always tested using the split of the data that was not used during training. We used a two-fold cross-validation procedure due to the excessive time taken to train the networks. To ensure that cross-validation was representative of the data, despite the small number of folds, the distribution of split-half reliabilities of all possible splits was computed and the split that best represented the mean of the distribution was chosen for all further analyses.

**Architectural overview**

Each network contains six convolutional layers followed by a linear readout. All convolutions have a stride of 1×1 and are padded such that the convolution leaves the height and width dimensions of the layer unchanged. Prior to each convolutional layer (except the first), the feedforward input to the network goes through a max pooling layer with 2×2 stride and a 2×2 kernel. This has the effect of reducing the height and width dimensions of the input by a factor of 2.

Architectural parameters are outlined in Supplementary Table 1 including the number of feature maps, kernel size and image size for each layer. The addition of lateral and top-down connections in BLT leads to an increased number of parameters compared to a feedforward B model. A larger kernel size is used in B to approximately match the number of parameters in BLT, while maintaining the same number of units and layers across the networks. As it is not possible to exactly match the number of parameters by adjusting the kernel size, we use the two closest B models, with kernel sizes of 9 and 11, subsequently referred to as $B_{K9}$ and $B_{K11}$, respectively.

For architectural simplicity, the kernel size was kept fixed throughout the networks. If the image size reduces to less than $(k + 1)/2$ (where $k$ is an odd kernel size), then the whole kernel is not used after it has been centred on each of the inputs. This reduces the effective kernel size for the layer, which only occurs in the final convolutional layer of B (see Supplementary Table 1). Taking the effective kernel size into account, the number of parameters sums to 3.0 million in $B_{K9}$, 4.3 million in $B_{K11}$ and 4.0 million in BLT. Time is modelled in the neural networks by defining each convolution as taking a single time step. In practice, it is easier to implement the feedforward connections as instantaneous, lateral connections as taking one time step and top-down connections as taking two time steps. These two definitions are computationally equivalent if lateral and top-down connections have no influence prior to the arrival of feedforward input to the layer.

**Recurrent convolutional layers**

The recurrent convolutional layer (RCL) forms the basis of the models used in these experiments. The activation in a single RCL is represented by the three-dimensional array $H_{\tau,n}$, the index $\tau$ is used to indicate the time step and $n$ is used to indicate the layer. The dimensions in $H_{\tau,n}$ correspond to the height, width and features in the layer. We define $H_{\tau,0}$ to be the input image to the network.

Convolutional weights for a given layer in the network are represented by the arrays $W_n$. All instances of $W_n$ are implemented using weight normalisation to assist learning (57). The biases for each layer are represented by the vector $\boldsymbol{b}_n$, with a unique bias for each feature map in the output.

For classic feedforward (B) networks, the lack of recurrent connections reduces RCLs to a standard convolutional layer

$$H_{\tau,n} = \left[W_n^b * H_{\tau-1,n-1} + \boldsymbol{b}_n\right]_+$$

Where $W_n^b$ represents the bottom-up convolutional weights and $[\cdot]_+$ is the rectified linear function. All layers are made inactive prior to the arrival of feedforward input to the layer by defining $H_{\tau,n} = 0$ when $\tau < n$.

As standard feedforward networks lack dynamics, we modify the B layers to allow units to ramp-up their activation over time via self-connections. Self-connection weights are controlled by the parameter $\omega_n$, which is shared across the layer. Note that this model class contains conventional feedforward models as a special case, where $\omega_n = 0$. The self-connection weights were constrained to be nonnegative and optimized along with the other connection weights.

$$H_{\tau,n} = \left[ W_n^b * H_{\tau-1,n-1} + \omega_n H_{\tau-1,n} + b_n \right]_+$$

BLT layers are formed by the addition of lateral and top-down convolutions with weights $W_n^l$ and $W_n^t$, respectively.

$$H_{\tau,n} = \left[ W_n^b * H_{\tau-1,n-1} + W_n^l * H_{\tau-1,n} + W_n^t * H_{\tau-1,n+1} + b_n \right]_+$$

Max-pooling has the effect of reducing the height and width dimensions of RCLs as we move up the layers of the network. This means that the size of the outputs from top-down convolutions does not match the size of the outputs for bottom-up and lateral convolutions, as the convolutions preserve image size. To compensate for this, we apply nearest-neighbor up-sampling to the output of the top-down convolution to make the sizes match. This has the effect of small, non-overlapping patches of neighboring units receiving identical top-down input.

In the final BLT layer, top-down input is drawn from the readout layer of the network. In this case, a fully connected layer is used for top-down connections as opposed to the convolutional layer that is used elsewhere.

**Readout layer**

A linear readout is added to the end of the network to produce an output, $h_{\tau,\text{cat}}$, for each of categories that the network is trained on.

Prior to the readout, the bottom-up input goes through global average pooling. This averages over the spatial dimensions of final layer, $N$, to produce a vector with length equal to the number of features in the final layer, which we denote $\bar{h}_{\tau-1,N}$.

The readout layer is also provided with lateral input from the readout on the previous time step, $h_{\tau-1,\text{cat}}$. This allows the network to sustain categorisation responses without depending on continuous bottom-up input.

In B networks, lateral inputs take the form of self-connections that enable the units to increase their activation over time, in the same manner as the B convolutional layers.

$$h_{\tau,\text{cat}} = W_{\text{cat}}^b \bar{h}_{\tau-1,N} + \omega_{\text{cat}} h_{\tau-1,\text{cat}} + b_{\text{cat}}$$

Where $W_{\text{cat}}^b$ are fully-connected bottom-up weights.

In BLT networks, the readout units have a set of fully connected lateral weights, $W_{\text{cat}}^l$, so each readout unit receives input from all other readout units.

$$\boldsymbol{h}_{\tau,\text{cat}} = W_{\text{cat}}^b \, \overline{\boldsymbol{h}}_{\tau-1,N} + W_{\text{cat}}^l \, \boldsymbol{h}_{\tau-1,\text{cat}} + \boldsymbol{b}_{\text{cat}}$$

**Training**

The networks were trained using a two objectives, representational distance learning and object classification.

*Representational distance learning*

We extended representational distance learning(36) to be used as an objective which aims to match network representational dynamics across multiple selected layers to the RDM movies of three ventral stream regions. Input images were taken from the RDL61 image set, which matches the categorical structure of the experimental stimuli. We use RDL to train layers 2, 4 and 6 of the network to match the dynamics of V1-3, V4t/LO, and IT/PHC, respectively.

The ventral stream RDMs undergo several pre-processing steps before being used for RDL. First, distances are averaged across any of the 92 images that fall into the same category in RDL61. For instance, the 92 stimuli contain 12 images of faces that constitute a single category in RDL61. Since optimization was performed at the category level, a single distance estimate was obtained as the average across all face distances. Averaging distances over categories produces a 61×61 RDM for each time point in the MEG data. Each of the reduced RDMs are down-sampled from 500Hz to 200Hz by taking average RDMs over 5ms time windows centred at 5ms intervals from $t_{\text{start}}$ to $t_{\text{start}} + 250\text{ms}$. The value of $t_{start}$ varies for each of the ROIs, for V1-3 $t_{\text{start}} = 50\text{ms}$, for V4t/LO $t_{\text{start}} = 60\text{ms}$, and for IT/PHC $t_{\text{start}} = 70\text{ms}$. The delay between each of the ROIs was used to account for the time taken to perform feedforward processing, as the model does not process information prior to arrival of feedforward input.

To apply RDL, mini-batches are divided into $M/2$ pairs, where $M$ is the batch size. Images in the mini-batch are pseudo-randomly sampled so that each pair contains two images, $x_a$ and $x_b$, from two different categories, category $a$ and category $b$. Within a pair we calculate the correlation distance between the network activations in a given layer in response to these two images, $\hat{d}_{\tau,n}(x_a, x_b)$. This was performed for each time point and layer where RDL is applied. To compute the error for RDL, we compare $\hat{d}_{\tau,n}(x_a, x_b)$ to the distance for the two categories in the ventral stream MEG data, $d_{\tau,r}(a,b)$.

$$E_{\text{RDL}} = \sum_{n \in L, r \in R} \frac{1}{T} \sum_{\tau}^{T} \frac{\left(\hat{d}_{\tau,n}(x_a, x_b) - d_{\tau,r}(a,b)\right)^2}{\sigma_{\tau,r}^2}$$

Where $L = \{2, 4, 6\}$ represents the network layers where RDL is applied and $r$ represents the corresponding ROI from the set of all ROIs, $R$, that were used in training. We use the variance of the empirical RDM at each time step, $\sigma^2_{\tau,r}$, as a normalisation factor. This normalisation prevents the loss from being biased towards time points with larger variance in the RDMs. As a result, each time-point will impact the optimization independently of the RDM variance/noise level.

*Categorization objective*

The loss for categorization is calculated in two stages. Firstly, the softmax output $\hat{y}_{\tau,i}$ is computed from the readout layer of the network for every category and time point. The error for the categorization objective, $E_{\text{cat}}$, is computed by calculating the cross-entropy between the softmax output and target for each category output $y_i$ (where the target category is represented with one-hot encoding) and then averaging across time.

$$E_{\text{cat}} = -\frac{1}{T}\sum_{t=1}^{T}\sum_{i=1}^{C} y_i \cdot \log \hat{y}_{\tau,i}$$

Where $C$ represents the number of categories used during training and $T$ is the total number of time steps.

*Overall objective*

A combination of the RDL and categorization objectives, with additional L2-regularisation, produces the overall loss function for the network.

$$\mathcal{L} = \gamma_{\text{cat}}\bar{E}_{\text{cat}} + \gamma_{\text{RDL}}\bar{E}_{\text{RDL}} + \lambda|w|_2$$

Where $\bar{E}_{\text{cat}}$ and $\bar{E}_{\text{RDL}}$ are the average of $E_{\text{cat}}$ and $E_{\text{RDL}}$ over the mini-batch. The contribution of each objective is controlled by the two coefficients $\gamma_{\text{cat}}$ and $\gamma_{\text{RDL}}$. The level of L2-regularization is controlled by the coefficient $\lambda = 10^{-5}$, and $w$ represents all weights of the network in vectorised format.

We set $\gamma_{\text{RDL}} = 1$ throughout training, and initially set, $\gamma_{\text{cat}} = 10$, which causes the categorization loss to dominate at the beginning of training (Figure 3A). Over the course of training $\gamma_{\text{cat}}$ decays by a factor of 10 every 10,000 mini-batches until it reaches a value of $10^{-2}$, where it remains constant until training terminates after 4 million mini-batches.

We use Adam (58) to optimize the overall loss with the following parameters, learning rate $\alpha = 10^{-3}$, exponential decay parameters $\beta_1 = 0.9$ and $\beta_2 = 0.999$, and stabilisation parameter $\hat{\epsilon} = 10^{-1}$. See Supplementary Figure 7 for image classification test performance across training for the different model types.

**Virtual cooling studies**

To emulate cortical cooling studies, we used dropout at different keep probabilities to specifically target lateral and top-down connections in the computational graph of the network. Dropout was applied independently to the output of the lateral and top-down convolutions. The change in the mean activity level of the network/layer was corrected. The resulting network activations were then tested for (a) their ability to predict the representational dynamics observed in the human ventral stream, and (b) their ability to perform object categorization. To assess the ability to predict human ventral stream data, we computed the average correlation of the frames of the network RDM predictions and the empirical RDM movies (similar to Figure 3C). To assess the networks' ability to perform object categorization, we computed accuracy using the validation set (the accuracy metric was weighted such that each recognition performance for each class contributed equally to the overall score).

**Model fitting for off-the-shelf architectures**

Off-the-shelf feedforward DNNs trained for classification have shown early success in predicting time-averaged neural response profiles. Despite providing static output, as each layer produced only one activation vector at its output, the predictive performance of these models in the current dynamic setting can be informative.

As candidate models, we tested Alexnet (59) and VGG16 (60), which are both used frequently in visual computational neuroscience. To select the best layer for a given ventral stream ROI, we probed the networks with the same image dataset (RDL61) used previously for training our recurrent models. The networks were shown sets of randomly sampled 61 images, one for each visual category used in the human imaging experiments. The corresponding activation vectors (taken post ReLU) were then used to compute layer-specific RDMs. These RDMs were compared against the empirical data and the layer best predicting the whole timeseries of a given ROI was stored. This process was repeated 1000 times. Finally, for each ROI the network layer that was most frequently selected during bootstrapping was used for follow-up tests. Following the same cross-validation procedures used for RDL training above (session split half of the data, layer selection based on one half, test for predictive performance on the other half), we then extracted the RDMs from the winning layers for the 92 experimental stimuli. These RDMs were transformed to static RDM movies and subsequently entered the same analysis pipeline as the RDL networks (BLT, $B_{K9}$, and $B_{K11}$).

The Alexnet layers best predicting the MEG data were identical for the two cross-validation splits. Layers L5, L2, and L2 were selected for V1-3, V4t/LO, and IT, respectively. For VGG16, the cross-validation selected layers were: 5 and 12 (relu3_1 and relu5_2) for the two splits of V1-3, respectively, and ReLU layer 13 (relu_5_3) for Vet/LO and IT/PHC.

**Acknowledgments**. This research was funded by the UK Medical Research Council (Programme MC-A060- 5PR20), by a European Research Council Starting Grant (ERC-2010-StG 261352), by the Human Brain Project (EU grant 604102), and a DFG research fellowship to TCK. We would like to thank Rogier Kievit and Martin Hebart for comments on an earlier version of this manuscript.

**Author Contributions**. Conceptualization: TCK, NK; Methodology: TCK, CS, OH, NK; Formal Analysis: TCK, CS, LS; Investigation: TCK, RC, CS, LS; Writing - Original Draft: TCK, CS, NK; Writing – Review and Editing: TCK, CS, LS, RC, OH, NK; Funding Acquisition: NK, RC, TCK, LS

**Competing interests.** The authors declare no competing interests.

**Supporting Information:**
Materials and Methods
Figures        S1-S9
Tables         S1
Movies         S1-S6

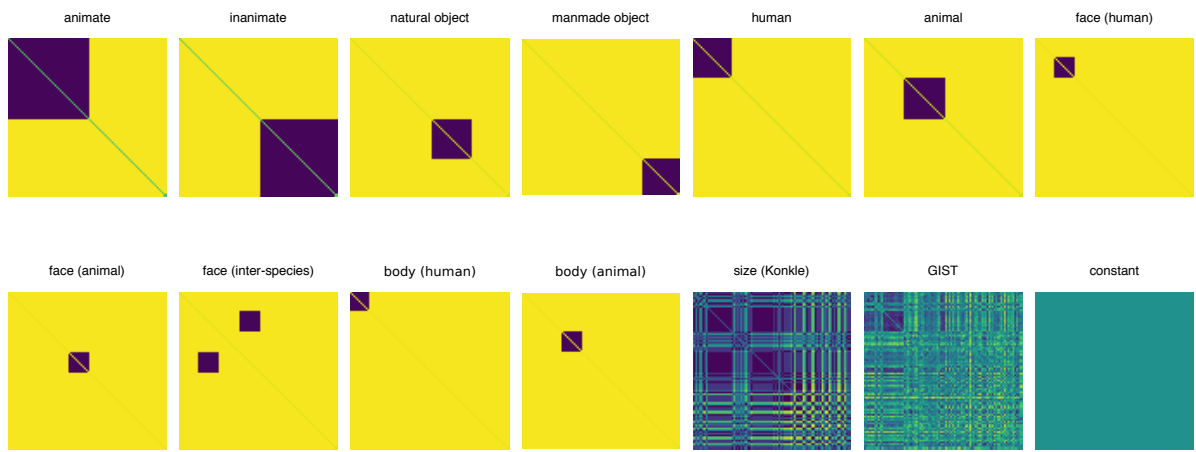

**Supplemental Figure 1. RDM components used as predictors during hierarchical GLM modelling.** The overall GLM included 11 categorical predictors, as well as a predictor derived from low-level Gist features, real-world size, and a constant.

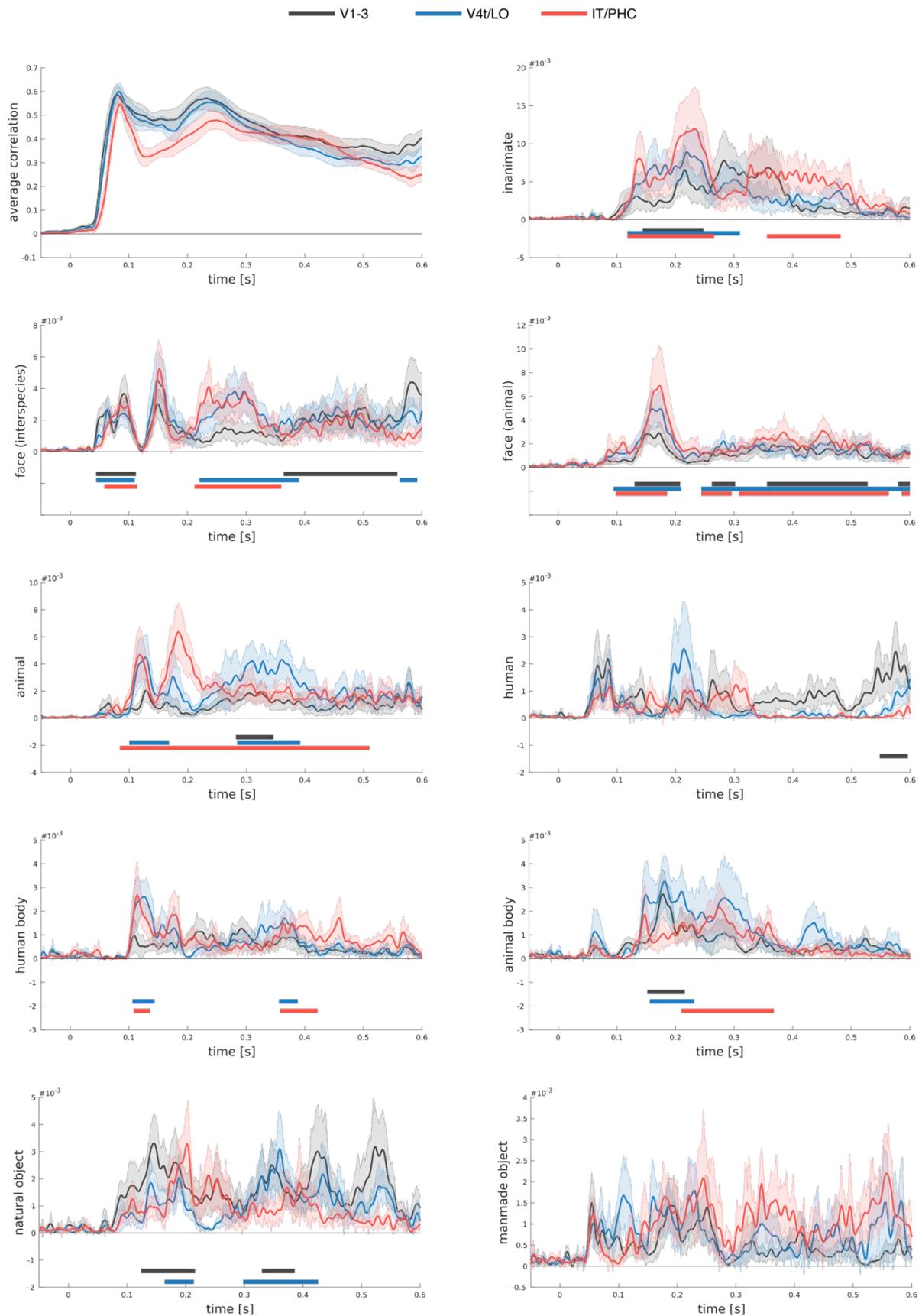

**Supplemental Figure 2. Supplemental results of the hierarchical GLM model fitting procedure.** Top left: average pattern distance across time for the three ventral stream ROIs. All others: unique variance explained by all control model predictors.

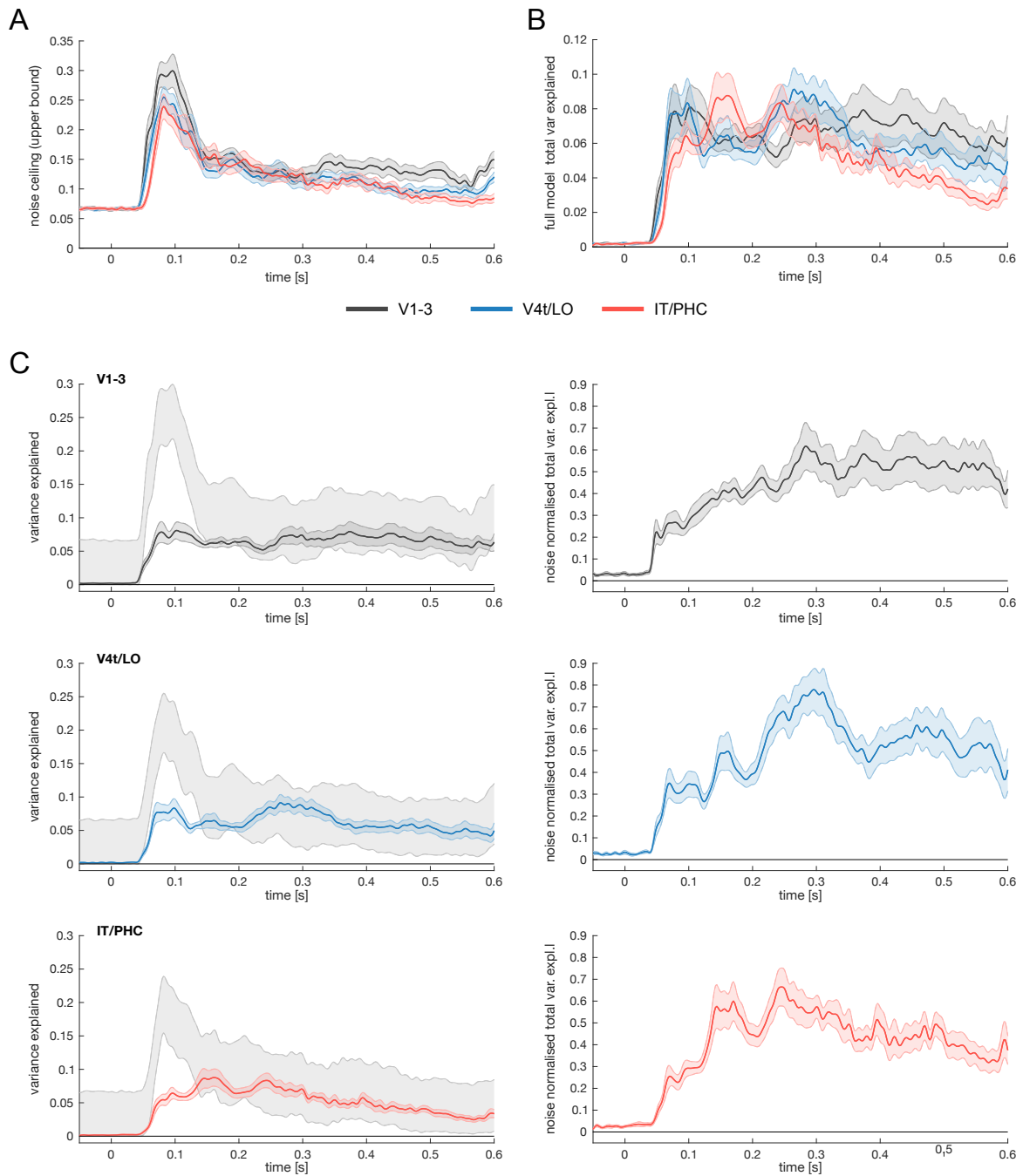

**Supplemental Figure 3. Noise ceiling estimates and total variance explained.** (A) Upper bound of the noise ceiling computed for each ventral stream ROI. (B) Total variance explained by the linear model used to dissect the ventral stream RDMs. (C) Upper and lower bound of the noise ceiling shown together with the total variance explained by the linear model (left column). The total variance explained was divided by the upper bound of the noise ceiling to express the percentage variance explained of the total explainable variance (right column).

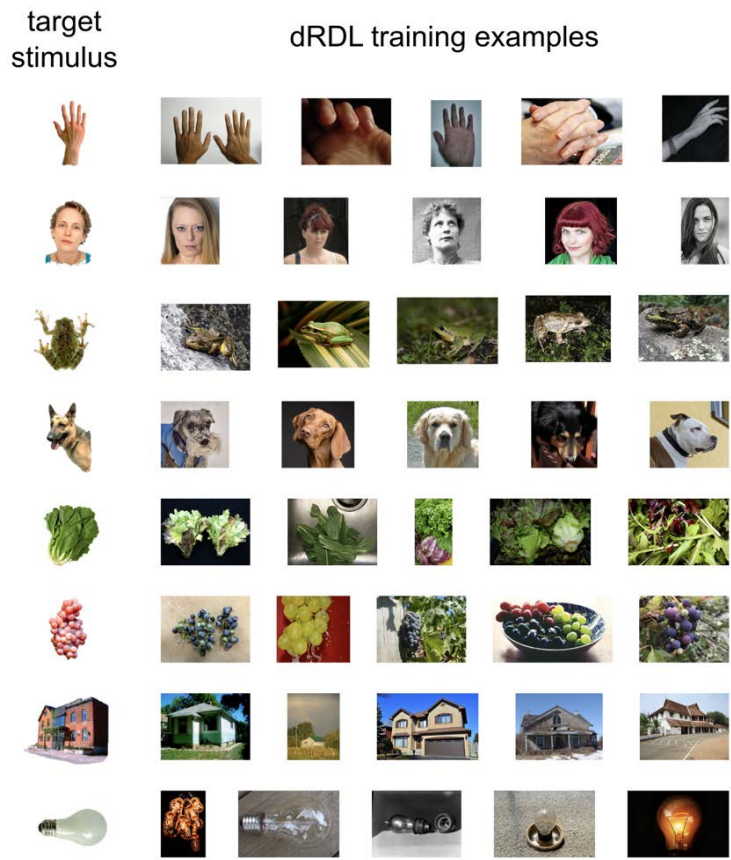

**Supplemental Figure 4. Example images used for RDL training.** The training set consisted of a total of 141k images, matching the categorical structure of the 92 experimental stimuli, while not including the original experimental stimuli. 61 categories were included for network training, directly mapping to 89 out of 92 experimental stimuli (3 stimuli were excluded from RDL training due to an insufficient number of images for training).

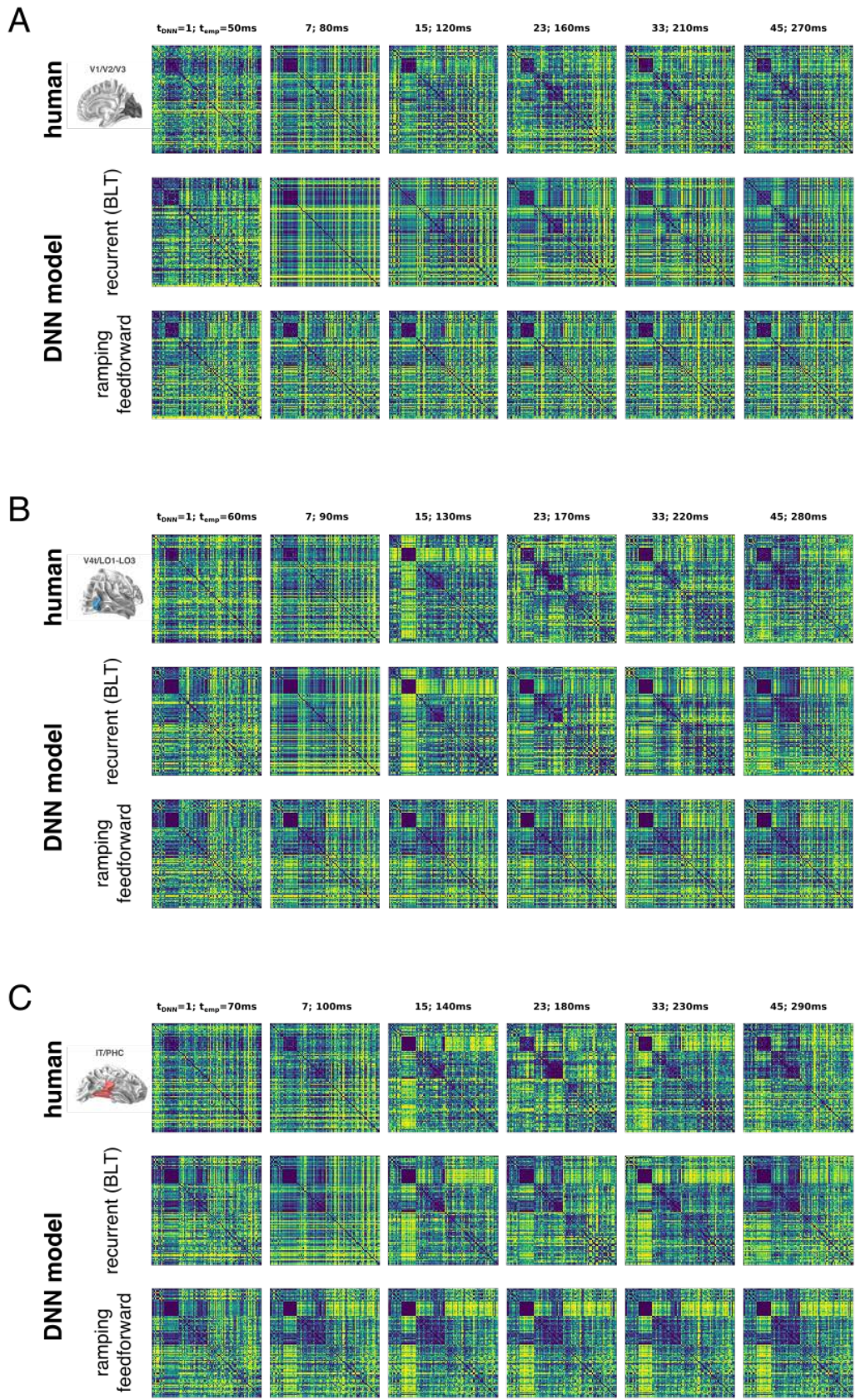

**Supplemental Figure 5. RDM movie frames for the representational trajectories of the human ventral stream ROIs and model architectures (ramping feedforward and recurrent (BLT) DNNs).** A-C show V1-3, V4t/LO, and IT/PHC, respectively. Numbers on top relate to the respective network and empirical brain times ($t_{DNN}$ and $t_{emp}$).

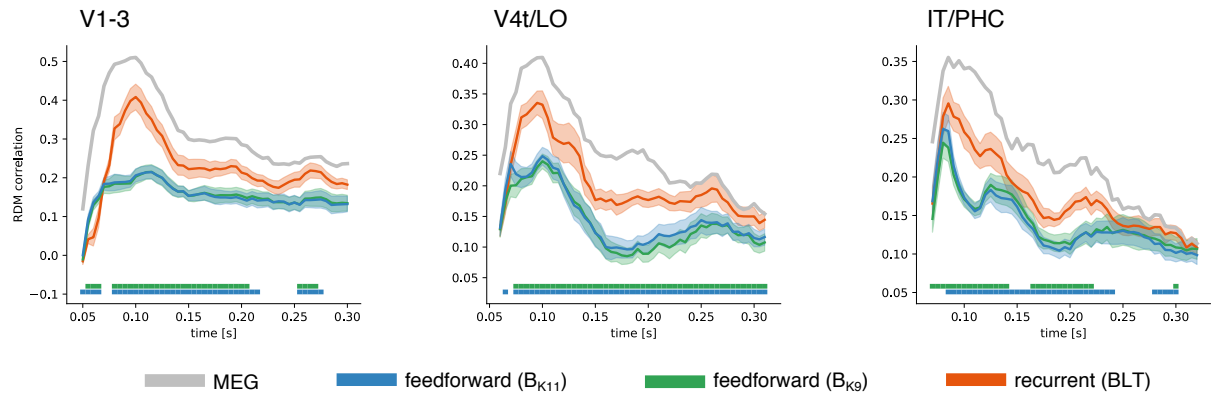

**Supplemental Figure 6. Frame by frame traces of RDM correlations between DNNs and brain data.** Split-half correlation for the MEG data, as used for training, shown in grey.

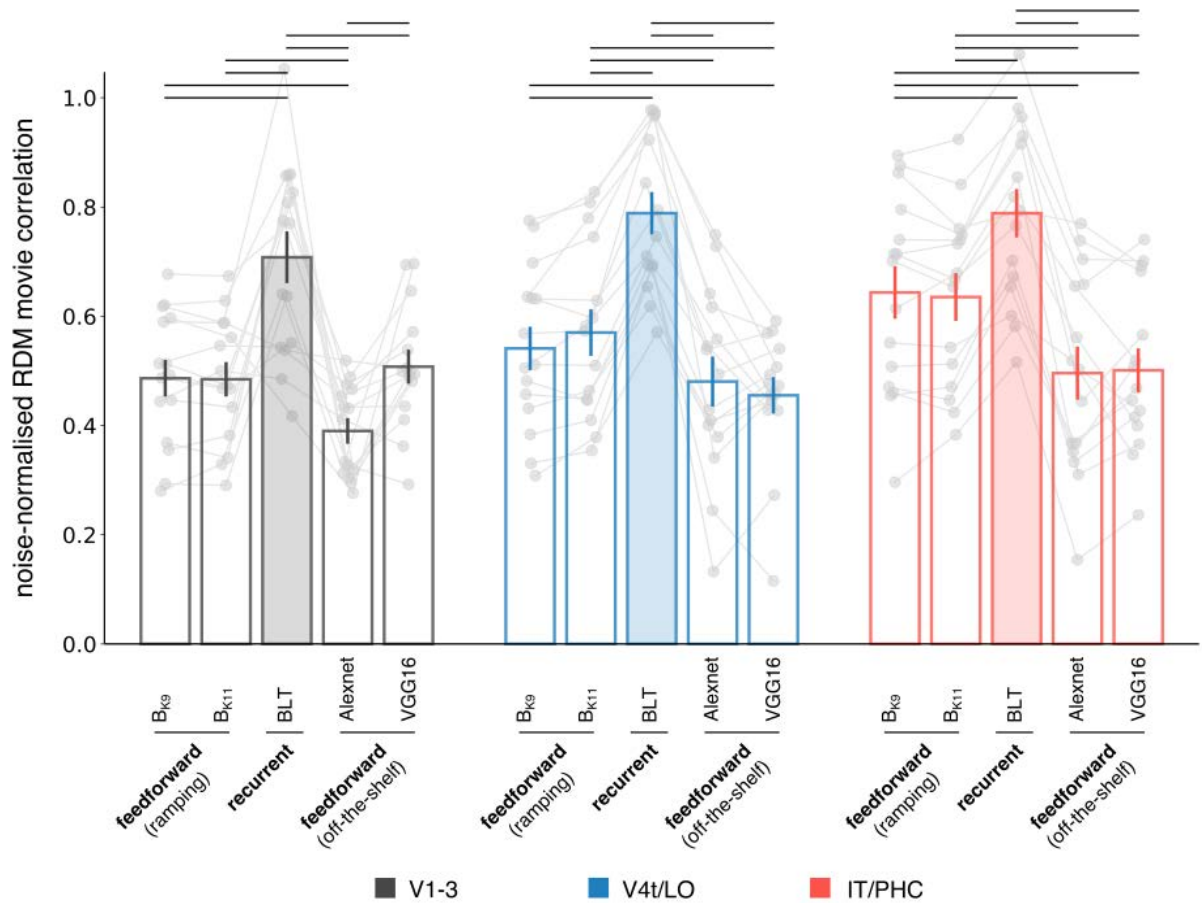

**Supplemental Figure 7. Comparison of our CNN models (ramping feedforward and recurrent) with off-the-shelf computer vision models Alexnet and VGG16.** Average frame-by-frame RDM correlation between deep network models and brain. Alexnet and VGG16 layers were selected via cross-validation. Contrary to the recurrent and ramping feedforward models, their RDM predictions are not time-varying, as classic feedforward models do not exhibit within-layer dynamics.

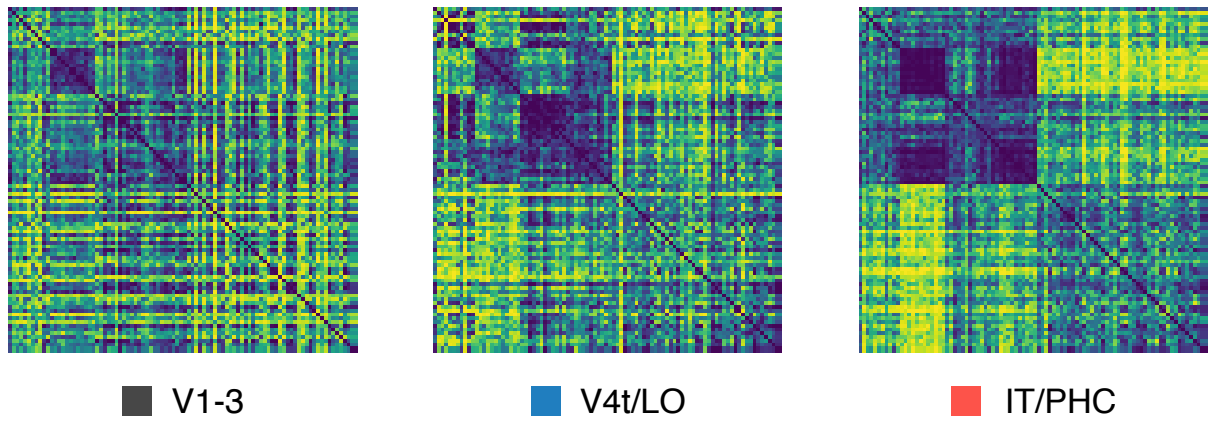

■ V1-3　　　　　　　　■ V4t/LO　　　　　　　　■ IT/PHC

**Supplemental Figure 8. fMRI analyses.** RDMs extracted from fMRI data obtained from the same participants and the same three ventral stream ROIs.

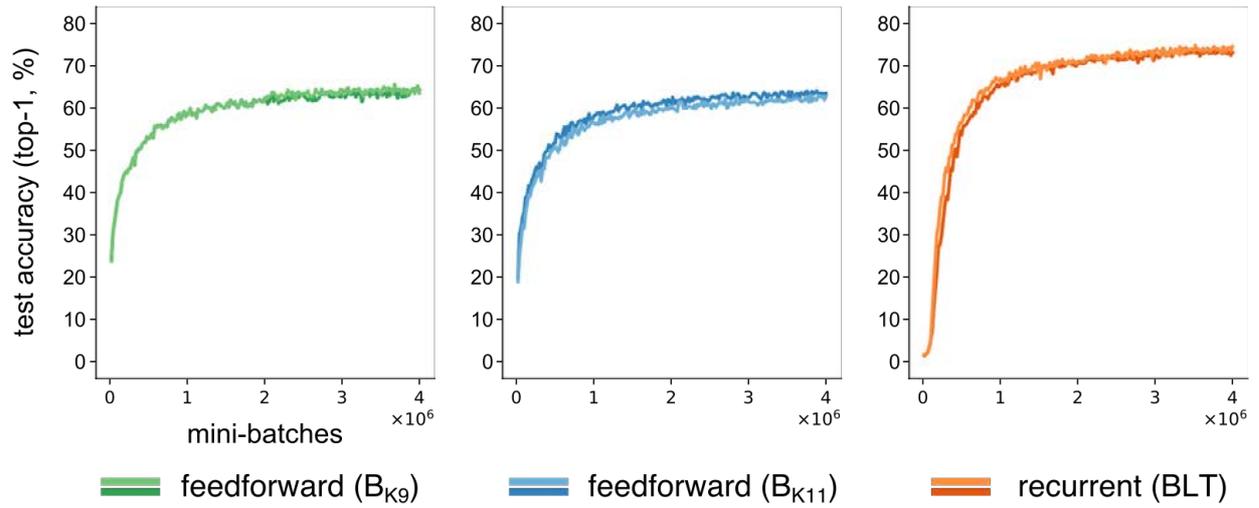

**Supplemental Figure 9. Image classification test performance across training for the different model types.** Ramping feed-forward ($B_{K9}$, $B_{K11}$; shown in green and blue, respectively) and recurrent (BLT; shown in orange) models.

**Table 1 Network architectural parameters**

| Layer | Feature maps | Image size | $B_{K9}$ – kernel size (effective) | $B_{K11}$ – kernel size (effective) | BLT – kernel size (effective) |
|---|---|---|---|---|---|
| 1 | 64 | 96 × 96 | 9 (9) | 11 (11) | 5 (5) |
| 2 | 64 | 48 × 48 | 9 (9) | 11 (11) | 5 (5) |
| 3 | 96 | 24 × 24 | 9 (9) | 11 (11) | 5 (5) |
| 4 | 96 | 12 × 12 | 9 (9) | 11 (11) | 5 (5) |
| 5 | 128 | 6 × 6 | 9 (9) | 11 (11) | 5 (5) |
| 6 | 128 | 3 × 3 | 9 (5) | 11 (5) | 5 (5) |

**Supplemental Table 1. Details on the network architectures.**

**Supplemental Movies**

1. **MEG RDM movies for three ventral stream ROIs.**
2. **Recurrent DNN RDM movies showing the reconstruction of the representational dynamics observed across three ventral stream ROIs.**
3. **Feedforward DNN RDM movies showing the reconstruction of the representational dynamics observed across three ventral stream ROIs.**
4. **MEG, feedforward and recurrent network RDM movies for area V1-3.**
5. **MEG, feedforward and recurrent network RDM movies for area V4t-LO**
6. **MEG, feedforward and recurrent network RDM movies for area IT/PHC**